Ph.D. Thesis

(Ph.D. thesis in basic of the dissertation)

# Mathematical Modeling of Aerodynamic "Space -to - Surface" Flight with Trajectory for Avoid Intercepting Process for Safety and Security Issues

Serge Gorneff

(Name updated)

(Space-Engineering Institute, Aerospace Cyber research Inc.)

## Abstract

Dissertation has been made for research project of mathematical modeling of aerospace system "Space-to-Surface" for avoid intercepting process by flight objects "Surface-to-Air". The dissertation includes Introduction, the 3 Chapters, and Conclusion. The research has been completed and created mathematical models which used for research and statistical analysis. In mathematical modeling has been including a few models: Model of atmosphere, Model of speed of sound, Model of flight head in space, Model of flight in atmosphere, Models of navigation and guidance, Model and statistical analysis of approximation of aerodynamic characteristics. Modeling has been created for a Space-to-Surface system defined for an optimal trajectory for targeting in terminal phase. The modeling includes models for simulation atmosphere, aerodynamic flight and navigation by an infrared system. The modeling simulation includes statistical analysis of the modeling results.

## Analysis and Introduction

The analysis of the current scientific results of mathematical modeling in aerospace field and analysis of requirements for software support in aerospace industry opening the research tasks for creating the models of aerodynamic flights.

The aerodynamic flight in atmosphere required safety and security issues for different humanitarian missions and civil actions. The flight in atmosphere with international standards must be secured and safety. According these requirements the research and software simulation in this dissertation created the new aerodynamic trajectory and technical parameters of flight for avoid intercepting process for safety and security.

In dissertation has been created the aerodynamic trajectory and found the parameters, mathematical algorithm for navigation for safety landing. The requirements for trajectory have been made for safety landing, with minimization of error of landing and with minimization of flight timeframe.

Analysis of research tasks initiated the creating models for avoid intercepting process for (different missions) and provide recommendations for aerospace industry.

In research work has been used the research tools and methods of mathematical modeling, systems analysis, software computer simulation, statistical analysis and used theory of the probability.

## Chapter I

Modeling has been made for a "Space-to-Surface" system for creating an optimal trajectory for targeting in terminal phase with avoids an intercepting process. The modeling includes models for simulation atmosphere, speed of sound, aerodynamic flight and the navigation by an infrared system. The modeling and simulation includes statistical analysis of the modeling results.

In dissertation has created modeling and simulation of aerodynamic flight. The flight has an optimal trajectory for targeting and may be used for research missions. [1-4]. The scenario, which uses the unmanned vehicle of the "Space-to-Surface" system with optimal trajectory, includes:

· Launching the vehicle from a space orbit with an altitude of (H=250-300km) by pulse from a space platform. In addition, the vehicle may be launched from an aircraft in the "Space-to-Surface" mission during space flight. · After making the pulse from the space platform the vehicle enters the atmosphere. In next phase, it maneuvers in the atmosphere taking on horizontal flight. While in horizontal flight the vehicle is searching for the target by the infrared guidance system. Automatic control of the flight has the following conditions of altitude:

Maneuver in atmosphere made with radius R=35-40km. ¨ Horizontal flight must have altitude H= 33-40 km for searching safety landing.

The time for space flight is T=15-20 min until the time the vehicle enters the atmosphere. The time T=15-20 min is needed for the flight from space to enter to atmosphere with angle of attack A=3-4°.

Trajectory of the vehicle in the atmosphere includes four phases:

1. Aerodynamic flight with the included entrance to the atmosphere with an angle of attack A=3-4°. This part of the trajectory has an altitude of H=90-100 km and a velocity of V=7.6 km/sec.

2. Maneuver in atmosphere from H=80-90 km to H=30-40 km in altitude. The velocity in this phase is V=5-7.6 km/sec.

3. Landing is searching while in horizontal flight. The velocity in this phase is V=3-5 km/sec.

4. The terminal phase, navigation and deployment of the landing process. The velocity in this phase is V=2.8-3 km/sec.

All phases of the flight simulation include automatic control of the flight using parameter U. The mathematical model includes differential equations with parameter U integrated in four phases of flight. In the first phase it has been used for was gravitational flight:

1. $U = \cos\theta$, where θ- angle between horizontals and vector of vehicle velocity.

In second phase it has been used for the maneuver in atmosphere with radius R=45 km:

2. $U = V^2/gR + \cos\theta$, where V- velocity of vehicle.

In third phase it has been used for when the flight takes on horizontal flight with searching landing:

3. $U = k(H-Y) + \cos\theta$, where k – proportional coefficient, H=35 km, Y –current altitude.

In fourth terminal phase it has been used for the infrared guidance to the landing:

4. U = k□, where □ - angle between vehicle and landing place.

In terminal phase the design of vehicle may include the engine and make accelerating of velocity on final phase of flight. The vehicle in terminal phase may make maneuvers for avoid intercepting processing. The time for flight in the atmosphere with an altitude of H=100 km to deploy the landing is T=60-90 sec with a horizontal range of X=520-625 km. The vehicle, searching for the landing, starts in phase 3 by the infrared guidance system or may using the GPS for targeting. This simulation has been created by the integration of differential equations and includes modeling of the vehicle flight; searching, guidance, and automatic control and deploys the landing. As stated the modeling includes models of atmosphere, speed of sound, for automatic control of flight and for an infrared guidance. The simulation has been created in the programming language FORTRAN. The modeling includes statistical analysis of the modeling results. The simulation produced results with high probability of targeting inaccuracy (with an range error of R=8-10m). The simulation of the vehicle is used with a few parameters:

· Weight m=1450-1550 kg · Aerodynamic coefficient K=Cy/Cx~2 · Wing area S~2m²

Gravitational flight, maneuver in the atmosphere, horizontal flight, searching, guidance and deployment the landing are defined in the models differential equations. All parameters (X, Y, Z, V, T, U, A) of the flight are integrated, where:

· X, Y, Z- coordinates of vehicle; · T-time of flight; · V- velocity of vehicle; · U-parameter of automatic control; · A-angle of attack.

In modeling of atmosphere has been used the function the speed of sound Vs depends of altitude H, and made the linear approximation:

If altitude H>80 km, the speed of sound is Vs=272, 6 m/sec;

If altitude in 54 km<H<80 km, the Vs=330, 8-2*(H-54) m/sec;

If altitude is 45, 5 km<H<54 km, the Vs=330, 8 m/sec;

If altitude is 25 km<H<45, 5 km, the Vs=295, 1 + 1, 8(H-25) m/sec;

If altitude is 11 km <H < 25 km, the Vs=295, 1 m/sec;

If H < 11 km, the Vs = 340, 28-4, 1*H m/sec.

Aerodynamic parameter CX had approximation:

Cx = C(I,1)*M*M + C(I,2)*M + C(I,3),

Where M –Mach number, I = 1, 2, 3 and C(I) have:

| | | |
|---|---|---|
| C(1,1) = 1, 37 | C(2,1) = -6 | C(3,1) = 0, 01416 |
| C(1,2) = 0, 2 | C(2,2) = 12 | C(3,2) = -0, 16993 |
| C(1,3) = 0, 2 | C(2,3) = - 5 | C(3,3) = 0, 51679 |

The atmosphere as function of altitude:

p(H) = p*exp(-K*H)

The differential equations for vehicle “Space-to-Surface” have definition:

dV/dt =-Cx*p(H)*V*S*V/2*m*g – g*sinθ;

dθ /dt= g/V*(n – cosθ);

dZ/dt= -V cosθ sinW;

dX/dt=V cosθ cosW;

dY/dt=Vsinθ;

dn/dt=(U-n)/T;

dW/dt=gn/Vcosθ

Where,

V –current speed of vehicle;

θ- angle between horizontals and vector of vehicle velocity;

X, Y, Z- coordinates of vehicle;

g= 9, 080665m/sec2;

U-parameter of automatic control;

In mathematical modeling of flight vehicle have been used and integrated the differential equations:

Analysis modeling results after computer software simulation declared in horizontal range X= 615000m:

| T (sec) | V (m/sec) | θ | X (m) | H (m) | U |
|---|---|---|---|---|---|
| 0 | 7873 | -0,0442 | 0 | 84109 | 0 |
| 10 | 7872 | -0,0556 | 78617 | 74127 | 0 |
| 20 | 7870 | -0,0691 | 157160 | 69169 | 0 |
| 30 | 7862 | -0,0815 | 235596 | 63241 | 0 |
| 40 | 7834 | -0,0939 | 313791 | 56362 | 0 |
| 50 | 7743 | -0,1064 | 391361 | 48564 | 0 |
| 60 | 7451 | -0,1192 | 467111 | 39984 | 0 |
| 70 | 6756 | -0,1484 | 537382 | 31005 | -6,406 |
| 80 | 4225 | -0,3814 | 591596 | 17874 | -13,413 |
| 90 | 2041 | -0,9921 | 613258 | 3099 | -14,998 |
| 92 | 2198 | -1,0866 | 615016 | -0,4 | 0,3202 |

Analysis of modeling results declared that the navigation started at T=68 sec. on flight from attitude H=32850 m. The guidance of vehicle started at 82 sec of flight in atmosphere and H = 14750 m.

Computer software application simulation declared parameters of flight in horizontal range X= 675000m:

| T (sec) | V (m/sec) | θ | X (m) | H (m) | U |
|---|---|---|---|---|---|
| 0 | 7872 | -0,044 | 0 | 83000 | 0 |
| 10 | 7871 | -0,056 | 78617 | 74127 | 0 |
| 20 | 7870 | -0,069 | 157160 | 69169 | 0 |
| 30 | 7863 | -0,0815 | 235596 | 63242 | 0 |
| 40 | 7834 | -0,0941 | 313791 | 51363 | 0 |
| 50 | 7743 | -0,1064 | 391361 | 48564 | 0 |
| 60 | 7451 | -0,1192 | 467111 | 39984 | 0 |
| 70 | 6577 | -0,1301 | 537397 | 31197 | 0,7875 |
| 80 | 4758 | -0,1607 | 594200 | 23121 | -1,9578 |
| 90 | 2646 | -0,2602 | 630188 | 15791 | -2,4762 |
| 100 | 1808 | -0,3436 | 649341 | 9709 | 0,0482 |
| 110 | 2075 | -0,3624 | 669360 | 2067 | 2,1956 |
| 114 | 2045 | -0,3413 | 675008 | 0 | 0,9674 |

The navigation started at 85 sec of flight, from attitude 19360 m and time for guidance was period T=29 sec.

The software application execution declared parameters of flight in the horizontal range X=800000 m:

| T (sec) | V (m/sec) | θ | X (m) | H (m) | U |
|---|---|---|---|---|---|
| 0 | 7872 | -0,0442 | 0 | 84015 | 0 |
| 10 | 7871 | -0,0566 | 78617 | 74127 | 0 |
| 20 | 7870 | -0,0691 | 157160 | 69168 | 0 |
| 30 | 7862 | -0,0851 | 235596 | 63241 | 0 |
| 40 | 7834 | -0,0939 | 313791 | 56362 | 0 |
| 50 | 7743 | -0,1064 | 391361 | 48564 | 0 |
| 60 | 7451 | -0,1192 | 467111 | 39984 | 0 |
| 70 | 6579 | -0,0927 | 537426 | 31352 | 14,1606 |
| 80 | 5239 | -0,0194 | 596353 | 28511 | 5,4428 |
| 90 | 4210 | -0,0091 | 643254 | 28115 | 0,8792 |
| 100 | 3485 | -0,0254 | 681530 | 27547 | -0,2019 |
| 110 | 2859 | -0,0983 | 716263 | 25702 | -2,2937 |
| 120 | 2300 | -0,2226 | 741770 | 21722 | -1,7110 |
| 130 | 1774 | -0,3483 | 760898 | 16094 | -0,9868 |
| 140 | 1871 | -0,3976 | 777575 | 9550 | 0,0003 |
| 150 | 2026 | -0,3919 | 795566 | 1780 | 2,5917 |
| 153 | 2196 | -0,3743 | 799979,7 | -0,2 | 0,9562 |

The navigation of vehicle has been started since 98 sec of flight in altitude H = 27700 m and X = 674400 m. The time frame for guidance was T =55sec and the statistical error was E~8-10m.

The next software execution declared the parameters of flight for horizontal range X = 950000m. Software application has been run a few hundred times for get statistical definition of parameters:

| T (sec) | V (m/sec) | θ | X (m) | H (m) | U |
|---|---|---|---|---|---|
| 0 | 7867 | -0,0460 | 15717 | 87753 | 0 |
| 10 | 7869 | -0,0559 | 78577 | 82475 | 0 |
| 20 | 7872 | -0,0684 | 157118 | 77573 | 0 |
| 30 | 7874 | -0,0808 | 235623 | 71693 | 0 |
| 40 | 7871 | -0,0932 | 314048 | 64842 | 0 |
| 50 | 7850 | -0,1056 | 392276 | 57040 | 0 |
| 60 | 7763 | -0,1180 | 469930 | 48321 | 0 |
| 70 | 7763 | -0,1308 | 545682 | 38853 | 0 |
| 80 | 6430 | -0,0567 | 615272 | 30482 | 0 |
| 90 | 5098 | -0,0034 | 672682 | 28386 | 5,8346 |
| 100 | 4197 | -0,0615 | 718651 | 30047 | 0,9322 |
| 110 | 3705 | -0,0044 | 757932 | 31421 | 0,9193 |
| 120 | 3335 | -0,0071 | 793064 | 31231 | 0,8599 |
| 130 | 3022 | -0,0215 | 824799 | 30833 | 0,5107 |
| 140 | 2741 | -0,0715 | 853566 | 29632 | -1,3384 |
| 150 | 2442 | -0,1790 | 879315 | 26502 | -2,3507 |
| 160 | 2036 | -0,3102 | 901148 | 21067 | -1,7193 |
| 170 | 1771 | -0,4015 | 918290 | 14496 | 0,1053 |
| 180 | 1776 | -0,4493 | 934412 | 7180 | 0,8066 |
| 190 | 1711 | -0,4047 | 950004 | -2 | 0,9235 |

For distance X = 950000m the navigation has been started since 112 sec of flight with altitude H = 31300m. Timeframe for guidance was T = 78 sec and statistical error of landing was E ~6m.

The modeling results have been executed in software applications in variety of horizontal distance of flight. An error of landing has been determinate from the time of navigation. If timeframe for navigation will be 80-90 sec, the error of landing was minimum E = ~ 15 m.

Modeling and analysis of results declared that, if timeframe of navigation T nav < 15 sec, an error of landing increase more than E >15m.

In analysis of modeling results of vehicle found that the errors of landing depended from timeframe of navigation:

| | | | | | | |
|---|---|---|---|---|---|---|
| Max Error (m) | 35 | 30 | 25 | 20 | 18 | 15 |
| T nav (sec) | 12 | 28 | 55 | 78 | 86 | 97 |
| X (m) | 615000 | 675000 | 800000 | 950000 | 1000000 | 1100000 |

Analysis confirmed that decrease of landing error is function of the navigation time. If more time for navigation and distance of flight longer, the landing error is less.

## Chapter II

Second part of dissertation includes mathematical modeling and the computer simulation of flight objects “Surface – to- Air”. This complex of mathematical models is including:

Modeling of trajectory, mathematical modeling functionality of engine, created mathematical algorithm of guidance and modeling of navigation in final phase. Computer simulation has been done in software application which written in programming language FORTRAN.

In this chapter has been completed research for modeling and the software simulation of intercepting process. Created the mathematical modeling and software applications for definition the trajectory, the navigation, the guidance and intercepting process. The research analysis and the results of modeling have been done with statistical definition and systems analysis.

Aerodynamic parameter CX had approximation:

Cx = C(I,1)*M*M + C(I,2)*M + C(I,3),

Where M –Mach number, I = 1, 2, 3 and C(I) have:

| | | |
|---|---|---|
| C(1,1) = 2,85 | C(2,1) = -4,31 | C(3,1) = 0, 0143 |
| C(1,2) = -2.85 | C(2,2) = 8,62 | C(3,2) = -1,184 |
| C(1,3) = 1,31 | C(2,3) = -3,31 | C(3,3) = 2,07 |

The speed of sound Vs has linear approximation in conditions:

If  H< 11km, the Vs=340,28-4,1*H;

If 11km<H<25km, the Vs=295,1

In mathematical modeling of flight interceptors have been used and integrated the differential equations:

dV/dt = (R-Cx*p(H)*V*V*S/2*M)-g*sinθ;

dθ/dt= g/V*(n- cosθ);

dX/dt=V* cosθ;

dY/dt=V* sinθ;

dM/dt=-m;

dn/dt=(U-n)/T

Where, R- force of engine for interceptor;

U-parameter of automatic control;

V –current speed of interceptor;

θ- angle between horizontals and vector of interceptor velocity;

g-gravitation constant;

M- current weight;

m-change weight per second.

For differential equations with definition in space XYZ (3D) for interceptors we have additional equations:

dW/dt=g*n/V* cosθ;

dZ/dt=-V** cosθ*sinW.

Where, W- angle between horizontals and the vector of interceptor speed.

In modeling of interceptors used aerodynamic parameters for two types for interceptors. For first type of interceptor the maximum speed was in modeling V~1600m/sec; for second type of interceptor the maximum speed was V ~ 700m/sec.

After integration of differential equation, the modeling results of parameters for first type interceptor we have (3D):

| T(sec) | V(m/sec) | θ | X(m) | H(m) | M(kg) | Z(m) |
|---|---|---|---|---|---|---|
| 2 | 109,2 | 2,3210 | 947965,7 | 85,5 | 859 | 50,4 |
| 4 | 361,2 | 2,3214 | 947641,8 | 429,9 | 759 | 55,1 |
| 6 | 586,9 | 2,3212 | 946991,1 | 1124,9 | 659 | 64,6 |
| 8 | 808,5 | 2,3211 | 946037,5 | 2144,9 | 559 | 78,4 |
| 10 | 1056,6 | 2,3210 | 944768,7 | 3503,1 | 459 | 96,9 |
| 12 | 1362,8 | 2,3209 | 943124,6 | 5264,2 | 359 | 120,8 |
| 14 | 1625,9 | 2,3209 | 941110,1 | 7422,7 | 302,8 | 150,1 |
| 16 | 1423,2 | 2,3208 | 939320,6 | 9339,7 | 302,8 | 176,1 |
| 18 | 1213,0 | 2,3206 | 937765,0 | 11005,5 | 302,8 | 198,8 |
| 20 | 1073,1 | 2,3205 | 936371,4 | 12497,8 | 302,8 | 219,1 |
| 22 | 1009,4 | 2,3201 | 935099,9 | 13859,0 | 302,8 | 237,5 |
| 24 | 971,6 | 2,3182 | 933925,2 | 15116,2 | 302,8 | 254,6 |
| 26 | 892,6 | 2,3171 | 932831,7 | 16286,4 | 302,8 | 270,5 |
| 28 | 854,3 | 2,3170 | 931807,0 | 17382,6 | 302,8 | 285,4 |
| 30 | 828,1 | 2,3165 | 930841,5 | 18415,4 | 302,8 | 299,4 |
| 32 | 820,1 | 2,3160 | 929927,6 | 19392,7 | 302,8 | 312,7 |
| 34 | 815,4 | 2,3155 | 929059,8 | 20320,6 | 302,8 | 325,3 |
| 36 | 810,2 | 2,3150 | 928916,5 | 20473,8 | 302,8 | 327,4 |

For the second type of interceptor the modeling results after integration differential equations declare parameters (2D and with U-parameter of automatic control):

| T(sec) | V(m/sec) | θ | X(m) | H(m) | M(kg) | U |
|---|---|---|---|---|---|---|
| 2 | 89,1 | 3,0181 | 947985 | 14,3 | 800,6 | -2,149 |
| 4 | 562,1 | 3,0181 | 947299,8 | 86,5 | 616,5 | -3,974 |
| 6 | 721,2 | 2,9553 | 945897,9 | 290,7 | 543,8 | -4,477 |
| 8 | 640,4 | 2,8587 | 944616,0 | 593,2 | 528,8 | -3,884 |
| 10 | 573,8 | 2,7412 | 943784,1 | 880,4 | 513,8 | -3,223 |
| 12 | 520,4 | 2,5978 | 943266,7 | 1141,1 | 498,8 | -2,787 |
| 14 | 440,2 | 2,4688 | 942825,1 | 1448,0 | 483,7 | -2,376 |
| 16 | 333,5 | 2,3631 | 942424,1 | 1802,8 | 468,7 | -1,996 |
| 18 | 275,9 | 2,2788 | 942059,9 | 2191,9 | 453,7 | -1,652 |
| 20 | 267,4 | 2,2140 | 941727,4 | 2604,9 | 438,7 | -1,347 |
| 22 | 265,3 | 2,2034 | 941413,9 | 3028,6 | 423,7 | -0,591 |
| 24 | 263,7 | 2,2034 | 941101,4 | 3450,6 | 408,7 | -0,591 |
| 26 | 261,0 | 2,2034 | 940788,9 | 3870,4 | 393,6 | -0,591 |
| 28 | 259,6 | 2,2034 | 940481,6 | 4287,9 | 378,6 | -0,591 |
| 30 | 258,3 | 2,2034 | 940212,1 | 4654,1 | 365,4 | -0,591 |

Maneuvers can provide avoiding the interception process. In modeling has been used systems analysis of the automatic control of flight, determinate the zones of interceptors, the time frame for intercepting, speed of interceptors and radios of interceptor maneuvers. These issues can give opportunity for creating mathematical algorithm of automatic control of vehicle flight with maneuvers for avoids intercepting process.

In the terminal phase the vehicle made the maneuvers for avoid intercepting process. The maneuvers have been provided by changing the automatic control in the final phase of guidance. Capabilities of maneuvers have been deployed in zone of interception. This situation has condition:

$$Tv > Ti$$

Where, Tv- the time frame of vehicle flight,

Ti- the time frame for intercepting process

The maneuvers of vehicle have been made by kinetic energy of flight. The maneuvers of interceptor have been made by interceptor vehicle force. The process of maneuvers deployed by infrared navigation system, when can be registered the boost of interceptor engine. The signal of automatic control for maneuvers keeping the trajectory of vehicle with the potential minimum of radios maneuvers.

If these conditions deployed, the guidance of interceptors be in temporarily in break, and vehicle be safe at this timeframe in flight.

The conditions of maneuvers for vehicle and interceptors are:

$$Vv*Vv/g*Rv + \cos\theta v = Vi*Vi/g*Ri + \sin\theta i;$$

And we can write:

$$Rv = Vv*Vv/(Vi*Vi/g*Ri + \cos\theta i + \cos\theta v)*g$$

Approximately be:

$$Rv = Vv*Vv*Ri/Vi*Vi$$

For Vv = 2 km/sec, Vi = 1,8 km/sec, the Ri = 3km, and we can found the radios of maneuvers Rv ~3,7 km.

In aerodynamic flight of vehicle until terminal phase the automatic control was determinate:

$$U = k(H – Y);$$

If using U=Vv*Vv/g*Rv, the coefficient will be: K = (Vv*Vv/gRv)/(Hi-Y)

For vehicle speed Vv = 2000 m/sec, the radios of maneuvers R = 10 km, and attitude H = 33 km and current Y =30 km, the coefficient be: K = 0,01. In computer software simulation the coefficient has been variety in rate: 0,01 < K < 0,001.

Coefficient K from geometry proportions was: K = tgA, where A -angle of attack.

Modifications an automatic flight control has been provided by change of coefficient K. Automatic flight control has capabilities for maneuvers by changes of angle of attack.

Estimation of effectively using interceptor has been determinate in the probability of interception, which has been change in depended of the flight distance D and the vehicle speed Vv.

In modeling has been made definition of 3 zones of interception:

Zone 1: 0,2 km < D < 10 km; 0,015 km < H < 3 km

Zone 2: 10 km < D < 30 km; 3 km < H < 10 km

Zone 3: 30 km < D < 70 km; 10 km < H < 24 km.

Output of computer modeling has been creating parameters:

T – flight time of interceptor;

Vi – interceptor speed;

θ- angle between horizontals and vector of interceptor velocity;

X – distance of flight in horizontal 0X;

H – attitude of flight interceptor in 0Y;

M- current weight;

U - parameter of automatic control;

Differential equations has been integrated by method Runge- Kutta with T=0,02 sec. period.

Calibration of interception for flight objects has been simulated in software application.

The flight objects had linear trajectory for calibrate mathematical modeling of interception. For second type of interceptor we have the parameters and probability P of interception in modeling:

| P2 | H (m) | D (m) | V (m/sec) |
|---|---|---|---|
| 0,5 | 1000 | 1000 | 400 |
| 0,6 | 3000 | 2000 | 700 |
| 0,8 | 5000 | 3000 | 750 |
| 0,8 | 7000 | 4000 | 750 |
| 0,8 | 9000 | 5000 | 700 |
| 0,7 | 11000 | 6000 | 600 |
| 0,6 | 13000 | 8000 | 500 |
| 0,5 | 15000 | 10000 | 450 |
| 0,4 | 18000 | 12000 | 400 |

For the first type of interceptor we have the probability and parameters:

| P 1 | H (m) | D (m) | V (m/sec) |
|---|---|---|---|
| 0,5 | 1000 | 1000 | 500 |
| 0,6 | 3000 | 2000 | 1000 |
| 0,8 | 5000 | 3000 | 1500 |
| 0,9 | 7000 | 4000 | 1650 |
| 0,85 | 9000 | 6000 | 1600 |
| 0,8 | 11000 | 7000 | 1400 |
| 0,7 | 13000 | 8000 | 1200 |
| 0,7 | 15000 | 10000 | 1000 |
| 0,6 | 18000 | 12000 | 900 |
| 0,5 | 20000 | 14000 | 800 |

Zones, where the probability of interception have maximum, has been determinate in computer modeling results:

For first type;

4 km < H < 14 km;

2,5 km < D < 10 km;

And for second type of interceptor;

3,5 km < H < 11 km

2 km < D < 6 km

Increase of landing error was function of turbulent of atmosphere, and has been mathematically been modeling with increase number of noises. The noises were modeling with standard sub application (subroutine) GAUSS.

## Chapter III

In third part of dissertation has been made analysis of modeling results, systems analysis of software simulation and statistics analysis. The research activities include modeling of intercepting process, statistical analysis of intercepting, calculation of the probability of interception.

In third part has been calculate of probability of interception in real time processing with integrated the differential equations with interval T=0.02 sec. The software application executed and provided the calculation the probability of interception P ~ 0.2 in terminal phase. The statistical analysis has been proceeding for a few hundred executions of software application.

Analysis of modeling results and statistical definition has been declared that interception of vehicle has very small probability P =0,1. The time of flight in atmosphere for this probability must be T 90-100sec, and distance of flight X = 625-675 km.

If distance of flight was increase X =675 km – 1100 km, the vehicle can be intercepted. In this phase the speed of vehicle was 1,8 km/sec < Vv < 2,1 km/sec, and probability of interception was P=0,1-0,2. If speed of vehicle less than 1800 m/sec, we have probability of interception for both types:

| P2 | P1 | Vv (m/sec) |
|---|---|---|
| 0,3 | 0,4 | 1700 |
| 0,35 | 0,5 | 1600 |
| 0,4 | 0,55 | 1500 |
| 0,45 | 0,6 | 1400 |
| 0,5 | 0,65 | 1300 |
| 0,55 | 0,7 | 1200 |

Results of modeling is declared that the increase the distance of vehicle flight not reasonable, because with less speed, the probability of interception is increasing. The increase distance of flight declared that vehicle has more time for intercepting process. The mathematical modeling and computer software simulation declared:

| P1 | P2 | X (km) | Vv (m/sec) |
|---|---|---|---|
| 0,2 | 0,1 | 700 | 2000 |
| 0,25 | 0,15 | 800 | 1900 |
| 0,3 | 0,2 | 900 | 1800 |
| 0,45 | 0,3 | 1000 | 1700 |
| 0,55 | 0,4 | 1100 | 1600 |
| 0,6 | 0,45 | 1200 | 1500 |

The effective intercepting process has been in conditions:

For first type interceptor;

$$2{,}5 \text{ km} < D < 10 \text{ km}$$

$$4 \text{ km} < H < 14 \text{ km}, \ P = 0{,}6, \ \text{if } V_v < 1500 \text{ m/sec}$$

For second type of interceptor;

$$2 \text{ km} < D < 6 \text{ km}$$

$$3{,}5 \text{ km} < H < 11 \text{ km}, \ P = 0{,}5, \ \text{if } V_v < 1500 \text{ m/sec}$$

Launch of interceptor has been made in zone where will point of intercross of vehicle and interceptors. In the big distances D more than 10 km the probability of intercepting began less because the speed of $V_v > 1500$ m/sec and interceptor has technical limitation for maneuvers and interception.

The landing error is function of speed vehicle, and the distance of flight. If more speed $V_v > 1800$ m/sec, is difficult make correct landing and error of landing increased. If the distance of flight is increase, the vehicle spends more time in zone of interception and the probability of interception increase:

| Max Error (m) | P |
|---|---|
| 15 | 0,3 |
| 20 | 0,35 |
| 25 | 0,4 |
| 30 | 0,45 |
| 35 | 0,5 |

Analysis of modeling declared that the vehicle speed must be Vv > 1900 m/sec and timeframe for flight 70-90 sec. In these conditions possible avoid intercepting process, and probability of interception was less that P < 0,1.

## Conclusion and general scientific results

The simulation was calculated with two-dimensional and three-dimensional flight, design and analysis. The modeling and the simulation was created using various software applications with definition of all parameters. The trajectory which is includes four phases of flight may avoid any intercepting process. The guidance created for each part of flight and includes targeting in very short time in atmosphere, less than $T < 90$sec. In mathematical modeling created the algorithm for the guidance and intercepting process. The probability of intercepting has been executed in software applications and calculate the probability of intercepting $P \sim 0,1$. In conclusion, the modeling has statistical definitions and systems analysis of the modeling results with recommendations for the design of a vehicle in aerospace industry.

In the dissertation has been created the scientific results:

The new aerodynamic trajectory for safety and security landing of vehicle with minimization of error landing and the flight timeframe;

The new mathematical algorithm for navigation of landing vehicle with minimization of error of landing;

The mathematical modeling of flight which include navigation and guidance;

Created the computer software simulation (created software application in FORTRAN) for analysis of statistical results this made in a few hundred executions of software applications.

The mathematical modeling of interception process with software computer simulation;

The definition of probability of intercepting process;

The definition of technical parameters of vehicle with requirements for safety landing for avoids intercepting process.